\documentclass{ws-procs975x65}
\usepackage{graphicx}
\newcommand{\be}{\begin{equation}}
\newcommand{\ee}{\end{equation}}
\newcommand{\lb}[1]{\label{#1}}

\newcommand{\obs}[1]{[#1]_{{\tsty {\ssty obs}}}}
\newcommand{\bigobs}[1]{\left[#1\right]_{{\tsty {\ssty obs}}}}
\newcommand{\gen}[1]{#1_{\ssty i}}
\newcommand{\sty}{\scriptstyle}
\newcommand{\ssty}{\scriptscriptstyle}
\newcommand{\tsty}{\textstyle}

\newcommand{\etal}{{\it et al.\ }}

\newcommand{\dl}{d_{\ssty L}}
\newcommand{\da}{d_{\ssty A}}
\newcommand{\dg}{d_{\ssty G}}

\newcommand{\dz}{d_{\ssty Z}}

\begin{document}
\title{RADIAL DENSITY STATISTICS OF THE GALAXY DISTRIBUTION AND THE 
       LUMINOSITY FUNCTION}

\author{ALVARO S.\ IRIBARREM$^\ast$, MARCELO B.\ RIBEIRO$^\dag$
        and WILLIAM R.\ STOEGER$^\ddag$}
\address{$^\ast$Valongo Observatory, Federal University of Rio de
         Janeiro-UFRJ, Brazil\\
         $^\dag$Physics Institute, Federal University of Rio de
         Janeiro-UFRJ, Brazil; e-mail: mbr@if.ufrj.br\\
         $^\ddag$Vatican Observatory Group, Steward Observatory,
         University of Arizona, U.S.A.}

%


\bodymatter

The supernova cosmology project aimed at obtaining model independent
distances was a major step forward in the observational
validation of cosmological models. However, when one deals with galaxy
observations we are still unable to translate their redshift measurements 
into distance measures without assuming a cosmological model. This means that
observationally determined number densities, like the galaxy luminosity
function (LF), obtained with data derived from observed galaxy number
counts of redshift surveys, still require the assumption of a cosmological
model. The current astronomical practice in LF determination is to choose
the comoving distance and carry out all calculations only with this choice.
This implies analytical limitations as this methodology renders impossible
the possibility of developing consistent tests between theoretical
predictions of cosmological densities and their observations in its full
extent, because to do so requires the use of various distance measures
defined in cosmology in order to compare theory and observations. This
project aims at developing methods able to overcome such limitations. By
extracting the number counts from the LF results we are able to build
different observational densities with all distance measures and test the
underlying cosmological model by comparing with their respective
theoretical predictions, all that being done in a relativistic number
counting framework.

The various cosmological distance definitions are connected by the 
{\em reciprocity law} due to Etherington (1933; see also Ellis 1971, 2007),
\begin{equation}
\label{rec}
\dl = (1+z)^2 \da = (1+z) \; \dg,
\end{equation}
where $\dl$ is the {\em luminosity distance}, $\da$ is the {\em angular
diameter distance} and $\dg$ is the {\em effective distance}. A fourth
distance definition, the {\em redshift distance} $\dz$, can be obtained by
using the Hubble law as its defining expression (Ribeiro 2005).
Differential number densities $\gamma$ can be obtained by using such
distances as follows,
\be
\gen{\gamma}=\frac{1}{4 \pi (\gen{d})^2}\frac{dN}{d(\gen{d})}
=\frac{dN}{dz}\left\{ 4 \pi (\gen{d})^2 \; \frac{d(\gen{d})}{dz}\right\} ^{-1}
\label{gamma},
\ee
where $N$ is the {\em cumulative number counts} and the label $i$
corresponds to the distance definitions above ($i = {\sty A}$, ${\sty G}$,
${\sty L}$, ${\sty Z})$. The number density $\gamma^*$ can then be
obtained as,
\be
\gen{\gamma}^\ast=\frac{1}{\gen{V}} \int\limits_{\gen{V}} \gen{\gamma}
\; d\gen{V} = \frac{3 N}{4 \pi (\gen{d})^3}, \lb{gest}
\ee
where $\gen{V}$ is the spherical volume defined by the distance $\gen{d}$.
These are radial statistical tools that will be used to study the
observational galaxy distribution (Ribeiro 2005). They are composed of a
geometrical part, inside the brackets in eq. (\ref{gamma}) and $\gen{V}$
in eq.  (\ref{gest}), and of a number counts part, the {\em differential
number counts} $dN/dz$ in eq. (\ref{gamma}) and $N$ in eq. (\ref{gest}).
The geometrical part comes from the assumed cosmological model which is
inferred from other observations, like supernovae. The differential
number counts can be obtained theoretically using geometrical arguments
(Ellis 1971; Ribeiro \& Stoeger 2003), yielding,
\be
\frac{dN}{dz} = (\da)^2 \, d\omega_{\ssty 0} \, n \, (1+z) \, \frac{dy}{dz},
\ee
where $\omega_{\ssty 0}$ is the observed solid angle, $n$ is the number
density in proper volume and $y$ is the affine parameter of the past
null-geodesic connecting source and observer. $dN/dz$ can also be estimated
observationally, by means of the selection function $\psi$ derived from
the LF $\phi$ of a given galaxy redshift survey dataset,
\be
\lb{psi}
\psi = \int^{\infty}_{l_{lim}} \phi(l) \, dl,
\ee
where $l = L/L^*$, $L$ is the luminosity, $L^*$ is the characteristic
luminosity parameter and $l_{lim}$ the luminosity limit
of the observations. This procedure incurs in some biases originating
from the fact that galaxy morphology and evolution are not being taken
into account in first approximation, but has the advantage of being
promptly applicable to the published LF of any galaxy redshift survey.

Ribeiro \& Stoeger (2003) connected the theoretical differential number
counts $dN/dz$ and its observational counterpart $\obs{dN/dz}$ by means
of the completeness function $J$ as follow,
\be
\lb{Jdef}
\bigobs{\frac{dN}{dz}} = J(z) \, \frac{dN}{dz},
\ee
whereas $J(z)$ itself can be obtained for a specific dataset by means of,
\be
J(z) = \frac{\psi(z)}{n_{\ssty C}},
\ee
where $n_{\ssty C}$ is the number density in comoving volume.

Ribeiro (2005) showed that the redshift evolution of $\gen{\gamma}$ and
$\gen{\gamma^*}$ are in fact affected by the distance definition. For a
Einstein-de Sitter universe it was shown that all such quantities, except
those based on $\dg$, grow increasingly inhomogeneous, in an observational
sense, with the redshift. Albani \etal (2007) showed that the same result
holds for the standard Friedmann-Lema\^{i}tre-Robertson-Walker (FLRW)
cosmology. Fig.\ \ref{gammathplot} shows the theoretical predictions of
the differential densities in a FLRW cosmological model with
$\Omega_{m_0}=0.3$ and $\Omega_{\Lambda}=0.7$.

Such results do not contradict the Cosmological Principle, since they
were all obtained assuming a {\em spatially} homogeneous FLRW universe.
It means, however, that null-geodesics effects in the redshift measurements
should affect the {\em observed} distribution of the galaxies. Rangel Lemos
\& Ribeiro (2008) showed that if an {\em observational} distribution of
galaxies were homogeneous, the resulting spatial distribution could be
inhomogeneous.
\begin{twocolumn}
Source undercounting appears to be introduced in the estimation of the
observed differential number counts and it seems to enhance the
inhomogeneity of the distributions, as shown in Fig.\ \ref{gammaplot}
for the observational differential number densities calculated using the
LF data of Gabasch \etal (2004) for the FDF survey. Indeed, Fig.\
\ref{Jplot} shows the redshift evolution of the completeness
function $J(z)$ for the dataset of Gabasch \etal (2004) and we can
see how such undercounting affects the observed number density in
comoving volume. 
\vspace{-1cm}
\begin{figure}
\begin{center}
\includegraphics[scale=0.5,angle=270]{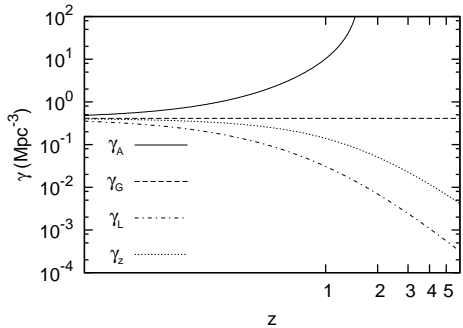}
\end{center}
\caption[Theoretical Differential Densities]
{ \small Redshift evolution of the theoretical differential densities $\gamma$
for the various cosmological definitions of distance. $\gamma_{\ssty G}$ is the
only one that remains homogeneous in the whole redshift interval. Ribeiro (2005)
showed that in the assumed FLRW cosmology this distance is equivalent to the
comoving distance and, therefore, $\gamma_{\ssty G}$ is homogeneous by
construction.
\label{gammathplot}}
\end{figure}
\begin{figure}
\begin{center}
\includegraphics[scale=0.5,angle=270]{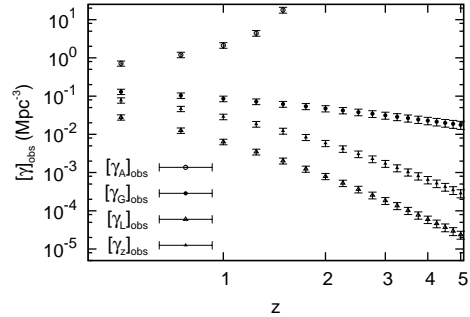}
\end{center}
\caption[Observational Differential Densities]
{\small Redshift evolution of the observational differential densities
$\obs{\gamma}$ for the various cosmological definitions of distance. Here
even $\obs{\gamma_{\ssty G}}$ gradually becomes inhomogeneous, which could,
perhaps, be credited to the effect of galaxy evolution and mergers in
the observed differential number count derived from the data. \label{gammaplot}}
\end{figure}
\vspace{-0.87cm}
\begin{figure}
\includegraphics[scale=0.5,angle=270]{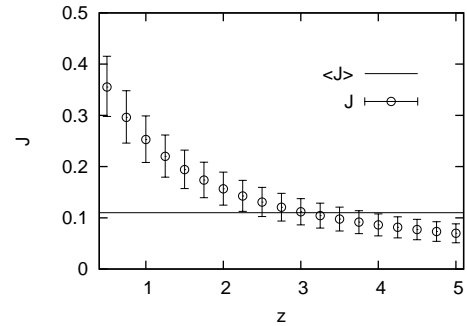}
\caption[Completeness function]
{\small Redshift evolution of the completeness function obtained from the
dataset of Gabasch \etal (2004). Clearly the discrepancy between
$\gamma_{\ssty G}$ and $\obs{\gamma_{\ssty G}}$ seems to stem from the
redshift evolution of $J(z)$ itself. The mean value of $J(z)$ in the
interval is plotted for comparison with its evolution.
\label{Jplot}}
\end{figure}

\end{twocolumn}

\end{document}